\documentclass[10pt, journal]{IEEEtran}
\IEEEoverridecommandlockouts

\usepackage{cite}
\usepackage{amsmath,amssymb,amsfonts}
\usepackage{algorithmic}
\usepackage{graphicx}
\usepackage{textcomp}
\usepackage{xcolor}
\usepackage{float}

\def\BibTeX{{\rm B\kern-.05em{\sc i\kern-.025em b}\kern-.08em
    T\kern-.1667em\lower.7ex\hbox{E}\kern-.125emX}}

\newcommand{\ov}[1]{\overline{#1}}

\begin{document}

\title{Local Synchronization of Power System Devices}

\author{Ignacio Ponce, \IEEEmembership{Student Member, IEEE}, and Federico
  Milano, \IEEEmembership{Fellow, IEEE}%
  \thanks{I.~Ponce and F.~Milano are with the School of Electrical and
    Electronic Engineering, University College Dublin, Belfield
    Campus, D04V1W8, Ireland. E-mails:
    \mbox{ignacio.poncearancibia@ucdconnect.ie},
    \mbox{federico.milano@ucd.ie}}%
  \thanks{This work is supported by the Sustainable Energy Authority
    of Ireland (SEAI) by funding I.~Ponce and F.~Milano under the project
    FRESLIPS, Grant No. RDD/00681.}%
  \vspace{-0.6cm} }%

\maketitle

\begin{abstract}
  This paper introduces a novel concept of local synchronization of power systems devices based on the difference between the complex frequency of the voltage and current injected at terminals.  Formal definitions are provided to account for bounded and asymptotic local synchronization.  The definitions are suitable for modern power systems as they remove classical assumptions limiting the application of the concept of synchronization to synchronous machines and omitting voltage dynamics.  The paper also provides a systematic analytical description of the synchronization mechanisms of common power system devices.  Finally, a variety of examples is included to illustrate the theoretical value and practical application of the proposed definitions to power systems modeling and stability analysis.
\end{abstract}


\section{Introduction}

In conventional power systems, the study of synchronization is substantially a synonym of transient stability analysis, the ultimate goal of which is to determine whether generators remain synchronized after a disturbance \cite{sauerpai}.  Specifically, the IEEE task force on the definition and classification of power system stability states: \textit{A machine keeps synchronism if the electromagnetic torque is equal and opposite to the mechanical torque delivered by the prime mover.  Accordingly, this type of stability depends on the ability of the synchronous machines to maintain or restore the equilibrium between these two opposing torques} \cite{taskforcestab}. 

The conventional understanding of synchronization no longer appears adequate for modern power systems.  It falls short of capturing the high heterogeneity of devices influencing the system's transient behavior and does not take into account the dynamics of the amplitude of the voltage.  For these reasons, in recent years, there has been a growing interest in revisiting the definition, interpretation and control of synchronization in power systems \cite{paganini, colombino, yangspectral, synchronizability, cfsyncdorfler, augmented}.  

Two common features of recently proposed definitions are that they focus only on the voltage at the point of connection of a device with the grid; and neglect the power exchange between the device and the grid.  This work proposes a general approach to define local synchronization based on the notion of complex frequency \cite{ComplexFreq} of voltage \textit{and} current at the point of connection of a device with the grid.

\subsection{Literature Review}

We review recent relevant works that have discussed the synchronization of power systems devices to the ac grid, specifically, power electronics converters.

In \cite{paganini}, the authors propose a model to study the global performance of synchronization in terms of industry-oriented metrics, such as nadir and RoCoF.  This approach incorporates heterogeneous models of synchronous machines and accounts for turbine dynamics. However, it neglects voltage amplitude dynamics and uses a highly simplified system model.  In \cite{colombino}, the authors recognize the role of the dynamics of the magnitude of the voltage in synchronization and propose a novel approach to study the \textit{global} synchronization of coupled oscillators founded on the seminal model of Kuramoto \cite{kuramoto}.  The goal is to study the stability of a specific control design for grid-forming converters, the dispatchable virtual oscillator control (dVOC) \cite{dVOC}.

In \cite{yangspectral}, the focus is on revealing the spectrum of the coupling matrix, identifying the conditions by which a negative eigenvalue arises, which possibly causes a loss of synchronism.  In \cite{synchronizability}, the authors recognize synchronization and stability are different, and develop a criterion for achieving synchronization in micro-grids with heterogeneous characteristics.  Particularly, \cite{synchronizability} aims at analyzing the effect of topology on \textit{synchronizability}, which they refer to as the robustness of synchronization against disturbances.  However, this definition depends on restrictive assumptions on the system model and is limited to phase synchronization only.

Other works incorporate the dynamics of the amplitude of the voltage by using the concept of the complex frequency (CF) \cite{cfsyncdorfler, augmented}.  The CF was recently formulated in \cite{ComplexFreq} to provide a more general and precise definition of the frequency in power systems, and how it is linked to the complex power variations.  Its formulation packs the full dynamics (in amplitude and phase) of dynamic vectors, vastly used to represent electrical quantities in power systems.  The CF is finding applications in modeling \cite{taxonomy, modalprop, cfdf, hybridcf} and control \cite{enhancing, virtualzcf, bernal2024improving} of modern power systems.  Despite remarkable advances made in \cite{cfsyncdorfler} using the CF to improve their previous stability analysis of dVOC grid-forming converters, it still limits the use of the concept of synchronization to study specific models in a grid composed entirely by converters with the aforementioned control scheme.

Another example of the use of CF to study synchronization is provided in \cite{augmented}, which removes simplifications made in previous synchronization studies that omit voltage dynamics and rely on network-reduced ODE models \cite{hillstabilitysync, motter2013spontaneous}. The paper translates classical transient stability analysis to the achievement of a state of ``augmented'' synchronization of bus voltages, namely, the convergence to a constant rate of change of magnitude and phase in all buses rather than the states of synchronous machines converging to an a-priori known equilibrium. The focus is on bridging the gap between practitioners, who empirically check the voltage and frequency trajectories to assess transient stability, and theorists, who rely on the Lyapunov-based theory.

\vspace{-2mm}
\subsection{Contributions}

We propose the novel concept of local synchronization. An important difference between our proposal and existing literature is that rather than a system-wide condition or property of the system, synchronization is assessed \textit{locally} for each device at the point of common coupling with the rest of the grid.  The proposed local synchronization definition aims at complementing the definition of the mathematical conditions for a proper power system operation.  A key feature of this definition is to be fully decoupled and independent from stability.  The proposed concept is practical as it is based on easily observable and measurable variables, that is, voltages and currents at the point of connection of a device with the grid.  Finally, the proposed device-oriented definition of synchronization is based on the concept of complex frequency and allows defining the conditions of synchronization in the form of explicit equations that depend on the internal states of the device itself.

The specific contributions of the paper are threefold, as follows.
\begin{itemize}
    \item A novel concept of local synchronization that allows discerning whether the dynamic behavior of a shunt-connected device is coherent with the rest of the system.  It is formalized through two model-agnostic definitions that are a function of electrical variables at the device terminal bus, namely, the difference between the CF of the current injected and the CF of the terminal voltage.
    \item A discussion on the difference between local synchronization and stability.  Even though synchronization and stability are commonly achieved (or not) together, we show that stability does not imply local synchronization, nor the other way around.  We also show that to reach desirable operating conditions following a transient, a power system has to be stable, but also every device has to achieve local synchronization.  The proposed concept brings completeness to the theoretical requirements for an adequate system operating condition.
    \item A description of the different synchronization mechanisms of a variety of common relevant devices.  This description is done systematically, and it is a result of combining the set of DAEs of the device model and the most restrictive of our definitions of local synchronization.  We show that, in general, device synchronization is subject to specific constraints on the CF of the terminal voltage to achieve synchronization.  For each device, we also identify the states involved in the synchronization mechanism and their underlying dynamic. 
\end{itemize}   

\vspace{-2mm}
\subsection{Organization}

The remainder of the paper is organized as follows.  Section \ref{sec:background} summarizes the theoretical background required to build the mathematical derivations and definitions presented in this paper. Section \ref{sec:localsync} formalizes the novel concept of local synchronization and proposes two definitions.  Section \ref{sec:taxonomy} describes the different synchronization mechanisms of common power system devices based on their dynamic model.  Section \ref{sec:examples} discusses several examples to evaluate local synchronization in a variety of scenarios using different devices through time domain simulations.  Conclusions and future work of the paper are given in Section \ref{sec:conclusion}.

\section{Background}\label{sec:background}
\subsection{Complex frequency}
The complex frequency (CF) of a Park vector $\bar{a}\in \mathbb{C}\,|\, \bar{a} = a\cos{\alpha}+\jmath\, a \sin{\alpha}$ is a complex quantity defined in \cite{ComplexFreq} as:
\begin{equation}
    \frac{\dot{a}}{a}+\jmath\,\dot{\alpha},\hspace{0.5 cm} a\neq 0 \, .
\end{equation}

Park vectors are used to model different electrical quantities, such as voltage $\bar{v}$, current $\bar{\imath}$, and admittance $\bar{y}$. The CF of those vectors is denoted as $\bar{\eta}$, $\bar{\xi}$ and $\bar{\chi}$, respectively.  Then, the time derivative of the Park vectors of voltage, current, and impedance can be written in terms of their CF \cite{ComplexFreq}:
\begin{equation}   
    \dot{\bar{v}} = \bar{v}\,\bar{\eta} \,,\hspace{0.5cm}  
    \dot{\bar{\imath}} = \bar{\imath}\, \bar{\xi}\,, \hspace{0.5cm}
    \dot{\bar{y}} = \bar{y}\, \bar{\chi}.
\end{equation}

\begin{figure}[hbtp]
  \centering
  \includegraphics[width = 0.5\columnwidth]{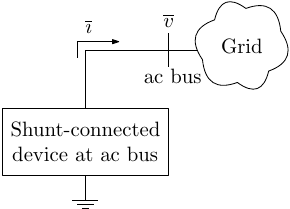}
  \caption{Shunt-connected device to the grid.}
  \label{fig:shuntdevice}
\end{figure}

Consider a three-phase balanced power system. The terminal voltage and current injected by a shunt-connected device at an ac bus are represented using Park vectors, denoted as $\bar{v},\bar{\imath}\in \mathbb{C}\,|\, ||\bar{v}||, ||\bar{\imath}|| \neq 0$, respectively.  Figure \ref{fig:shuntdevice} illustrates the situation.

It is always possible to calculate a \textit{dynamic equivalent admittance} of the device at terminals, $\bar{y} \in \mathbb{C}$, such that:
\begin{equation}\label{eq:iyv}
  \bar{\imath} = \bar{y}\, \bar{v} \, .
\end{equation}
The time derivative of (\ref{eq:iyv}) gives:
\begin{equation}
  \begin{aligned}
    \dot{\bar{\imath}} &= \dot{\bar{y}}\,\bar{v} + \bar{y} \, \dot{\bar{v}} \, , \\
    \bar{\imath} \, \bar{\xi} &= \bar{y}\, \bar{\chi}\,\bar{v} + \bar{y} \, \bar{v} \, \bar{\eta} \, , \\ 
    \bar{\imath} \, \bar{\xi} &= \bar{y}\, \bar{v} \,\left(\bar{\chi}+\bar{\eta}\right) \, , \\
    \Rightarrow \bar{\chi} &= \bar{\xi}-\bar{\eta} \, .
  \end{aligned}
\end{equation}
The CF of the dynamic equivalent admittance of the device, $\bar{\chi}$, is equal to the difference between the CF of the current injected at terminals and the CF of the terminal voltage. This paper discusses the local synchronization of devices in terms of the complex variable $\bar{\chi}$.

\subsection{General expression of $\bar{\chi}$}

In a previous paper \cite{cfdf}, we have shown that for shunt-connected devices, the complex frequency of the current injected at terminals, $\bar{\xi}$, can be written as:
\begin{equation}
  \label{eq:xi_dev}
  \bar{\xi}=\bar{\xi}_{\rm a}+\bar{\kappa}_\rho\rho+\bar{\kappa}_\omega\omega \, ,
\end{equation}
where $\rho,\omega \in \mathbb{R}$ are the real and imaginary parts of the CF of the voltage at the point of connection of the device, respectively.  Moreover, $\bar{\xi}_{\rm a}, \bar{\kappa}_\rho,\bar{\kappa}_\omega \in \mathbb{C}$ are complex quantities that (potentially) depend on (some of) the states and algebraic variables of the dynamic model of the device. 
Subtracting $\bar{\eta}$ from both sides of (\ref{eq:xi_dev}):
\begin{equation}
  \label{eq:chi_dev}
  \begin{aligned}
    \bar{\xi}-\bar{\eta}&=\bar{\xi}_{\rm a} +
    \bar{\kappa}_\rho\rho+\bar{\kappa}_\omega\omega-\bar{\eta}\\
    \bar{\chi}&=\bar{\xi}_{\rm a} +
    (\bar{\kappa}_\rho-1)\rho+(\bar{\kappa}_\omega-\jmath)\omega \, .
  \end{aligned}
\end{equation}

In the remainder of this paper, we use (\ref{eq:chi_dev}) to find an analytical expression for $\bar{\chi}$ to describe the different synchronization mechanisms devices have.

\section{Proposed Definition of Local Synchronization}
\label{sec:localsync}

Our starting point is to recognize that a shunt-connected device operates coherently if it can exchange active and/or reactive power with the rest of the grid.  In steady-state ac systems, this can happen only if the voltage and the injected current at the grid bus are isofrequential \cite{milano2020frequency}.  Using the framework of the complex frequency, this condition is fulfilled if the imaginary part of the CF of the voltage and current injected at terminals are equal, i.e., $\Im\{\bar{\xi}\}=\Im\{\bar{\eta}\}$ or, equivalently, $\Im\{\bar{\chi}\} = 0$.

It is also desirable to have an operation where $\Re\{\bar{\xi}\} = \Re\{\bar{\eta}\} = \Re\{\bar{\chi}\}=0$, which means an operation with a constant magnitude of the voltage and current injected at the terminals of the device.  In principle, a device could, for example, operate with steady-state oscillations in the magnitude of both variables.  In this case, $\Re\{\bar{\xi}\}$ and $\Re\{\bar{\eta}\}$ would be zero if averaged over a sufficiently long period.  A non-null average value is not sustainable in the long term and leads to the system collapse.

Based on these observations, we extend the desirable condition of a coherent operation of devices to the complex domain of frequency and to transient conditions.  We propose the following definitions. 

\vspace{3mm}

\textit{Definition 1}: A device achieves Bounded Local Synchronization (BLS) if the trajectory of $\bar{\chi}$ after the initial time $t_0>0$ stays bounded:
\begin{equation}
  \label{eq:sync_eps}
  \boxed{\forall \epsilon>0,\, \exists \delta >0\,|\, ||\bar{\chi}(t_0)||< \delta \Rightarrow ||\bar{\chi}(t)||< \epsilon, \,\forall t> t_0}
\end{equation}

\vspace{3mm}

\textit{Definition 2}: A device achieves Asymptotic Local Synchronization (ALS) if the trajectory of $\bar{\chi}$ tends to zero as time approaches infinity:
\begin{equation}
  \label{eq:sync_ss}
  \boxed{\lim_{t\rightarrow \infty}\bar{\chi}(t)=0}
\end{equation}

\vspace{3mm}

BLS accounts for the constantly varying nature of power systems which are always subject to small perturbations.  Thus, in the same vein stability is treated, small bounded variations of the trajectory of $\bar{\chi}$ around $0$ are acceptable to consider a device locally synchronous.  ALS represents the strict theoretical condition where an exact complex isofrequential condition is achieved.  Hence, ALS is a stricter condition than BLS.  When working with the set of equations composing the dynamic model of devices, the extreme case where $\bar{\chi}=0$ allows providing meaningful insights about the different synchronization mechanisms they have.  Several examples of this analytical development are presented in the following section.

In many practical situations, synchronization and stability are synonyms.  However, it is important to note that the proposed definition of local synchronization is independent and fully decoupled from the stability of the system.  unstable and devices not locally synchronous), Indeed, there are cases where the system is unstable but certain devices remain locally synchronous, and cases where the system is stable but a device is not locally synchronous.  These situations are illustrated in the case studies included in Section \ref{sec:examples}.

\section{Taxonomy of local synchronization for common power system devices}
\label{sec:taxonomy}

In this section, we present a systematic description of the synchronization mechanism of common power system devices.  This description requires the knowledge of the dynamic model of the device, i.e. its set of differential-algebraic equations (DAEs).  First, we present the set of DAEs and the expressions of the terms $\bar{\xi}_{\rm a}$, $\bar{\kappa}_\rho$ and $\bar{\kappa}_\omega$ that appear in \eqref{eq:xi_dev} for a variety of devices, ranging from conventional synchronous machines and loads to converter-interfaced generators.  The details of the analytical derivation required to get those expressions are not included to avoid unnecessarily overloading the manuscript with maths.  Detailed examples of this derivation can be found in \cite{cfdf}.  Finally, $\bar{\chi}$ is found using (\ref{eq:chi_dev}).  The resulting equations constitute an outcome of this work and allow describing the synchronization mechanism of each device in a general and systematic way.

\subsection{Synchronous machines}
\label{subsubsec:syn}

Consider Sauer Pai's 6th-order dynamic model of a synchronous machine \cite{sauerpai}, where symbols have the usual meanings:
\begin{equation}
\begin{aligned}
  \dot{\delta}&=\Omega_{\rm b}(\omega_{\rm r}-1)\label{eq:syn4_ddelta} \, ,\\
  M\dot{\omega}_{\rm r}&=\tau_{\rm m}-\tau_{\rm e}-D(\omega_{\rm r}-1) \, , \\
  T''_{\rm d0}\dot{\psi}''_{\rm d}&=-\psi''_{\rm d}+e'_{\rm q}-(x'_{\rm d}-x'_{\rm l})\imath_{\rm d}\, , \\
  T''_{\rm q0}\dot{\psi}''_{\rm q}&=-\psi''_{\rm q}-e'_{\rm d}-(x'_{\rm q}-x'_{\rm l})\imath_{\rm q}\, , \\
  T'_{\rm q0}\dot{e}'_{\rm d}&=(x_{\rm q}-x'_{\rm q})(\gamma_{\rm q1}\imath_{\rm q}-\gamma_{\rm q2}(\psi''_{\rm q}+e'_{\rm d})) -e'_{\rm d}\, , \\
  T'_{\rm d0}\dot{e}'_{\rm q}&=v_{\rm f}-(x_{\rm d}-x'_{\rm d})(\gamma_{\rm d1}\imath_{\rm d}-\gamma_{\rm d2}(\psi''_{\rm d}-e'_{\rm q}))-e'_{\rm q} \, , 
\end{aligned}
\end{equation}
where we have used the notation:
\begin{equation}
\begin{aligned}
    \gamma_{\rm d1}&=\frac{x''_{\rm d}-x_{\rm l}}{x'_{\rm d}-x_{\rm l}}\,, \quad \gamma_{\rm q1}=\frac{x''_{\rm q}-x_{\rm l}}{x'_{\rm q}-x_{\rm l}}\,,\\
    \gamma_{\rm d2}&=\frac{1-\gamma_{\rm d1}}{x'_{\rm d}-x_{\rm l}}\,, \quad \gamma_{\rm q2}=\frac{1-\gamma_{\rm q1}}{x'_{\rm q}-x_{\rm l}}\,,
\end{aligned}
\end{equation}
along with the algebraic equations that complete the model:
{\allowdisplaybreaks
\begin{align}
%\begin{aligned} 
\nonumber
  \psi_{\rm d} &=-x''_{\rm d}\imath_{\rm d}+\gamma_{\rm d1}e'_{\rm q}+(1-\gamma_{\rm d1})\psi''_{\rm d}\, ,\\ \nonumber
  \psi_{\rm q} &=-x''_{\rm q}\imath_{\rm q}-\gamma_{\rm q1}e'_{\rm d}+(1-\gamma_{\rm q1})\psi''_{\rm q}\, ,\\ \nonumber
  \bar{v} &= v_{\rm d} + \jmath \, v_{\rm q} = v\sin(\delta-\theta) +
            \jmath\, v\cos(\delta-\theta) \, ,\\
  \bar{\imath} &= \imath_{\rm d} + \jmath \, \imath_{\rm q} =
                 \imath\sin(\delta-\beta) + \jmath\,\imath\cos(\delta-\beta)
                 \, ,\\ \nonumber
  \bar{s} &= p + \jmath \, q = (v_{\rm d}\imath_{\rm d}+v_{\rm q}\imath_{\rm q}) +
            \jmath\, (v_{\rm q}\imath_{\rm d}-v_{\rm d}\imath_{\rm q}) \, ,\\ \nonumber
  \tau_{\rm e} &=\psi_{\rm q}\imath_{\rm d}-\psi_{\rm d}\imath_{\rm q} \, ,\\ \nonumber
  \bar{\psi} &= \psi_{\rm d}+\jmath\,\psi_{\rm q}=\jmath\, (R_{\rm s}\bar{\imath}+\bar{v})\, .
%\end{aligned}
\end{align}}
It can be shown that the results for $\bar{\xi}_{\rm a}$, $\bar{\kappa}_\rho$ and $\bar{\kappa}_\omega$ are as follows:
\begin{equation}
  \label{eq:syn6axikappa}
  \begin{aligned}
  \bar{\xi}_{\rm a} &= \jmath\,\omega_{\rm r} + \frac{\bar{\imath}^{*}\bar{Z}_{\rm d}(v_{\rm q}\omega_{\rm r}-\gamma_{\rm q1}\dot{e}'_{\rm d}+(1-\gamma_{\rm q1})\dot{\psi}''_{\rm q})}{(x''_{\rm d}x''_{\rm q}-R_{\rm s}^2)\imath^2}  \\
   &+ \frac{\jmath \,\bar{\imath}^{*}\bar{Z}^{*}_{\rm q} (v_{\rm d}\omega_{\rm r}+\gamma_{\rm d1}\dot{e}'_{\rm q}+(1-\gamma_{\rm d1})\dot{\psi}''_{\rm d}))}{(x''_{\rm d}x''_{\rm q}-R_{\rm s}^2)\imath^2} \, , \\
  \bar{\kappa}_{\rho} &= \frac{\bar{\imath}^{*}(\bar{Z}_{\rm d}v_{\rm d}- \jmath\, \bar{Z}^{*}_{\rm q} v_{\rm q})}{(x''_{\rm d}x''_{\rm q}-R_{\rm s}^2)\imath^2} \, , \\
  \bar{\kappa}_{\omega} &= \jmath\,\frac{\bar{\imath}^{*}(\jmath\, \bar{Z}_{\rm d} v_{\rm q} - \bar{Z}^{*}_{\rm q}v_{\rm d})}{(x''_{\rm d}x''_{\rm q}-R_{\rm s}^2)\imath^2} \, .
\end{aligned}
\end{equation}
We obtain the sought expression for $\bar{\chi}$:
\begin{equation}
  \label{eq:syn6chi}
    \boxed{\begin{aligned}
        \bar{\chi}=\jmath\,(\omega_{\rm r}-\omega)&\left(\frac{\bar{\imath}^{*}(\bar{Z}_{\rm q}^{*}v_{\rm d}-\jmath\,\bar{Z}_{\rm d}v_{\rm q})}{(x''_{\rm d}x''_{\rm q}-R_{\rm s}^2)\imath^2}+1\right) \\
        -\rho&\left(\frac{\bar{\imath}^{*}(\jmath\,\bar{Z}_{\rm q}^{*}v_{\rm q}-\bar{Z}_{\rm d}v_{\rm d})}{(x''_{\rm d}x''_{\rm q}-R_{\rm s}^2)\imath^2}+1\right) \\
        +&\frac{\jmath\,\bar{\imath}^{*}\bar{Z}_{\rm q}^{*}(\gamma_{\rm d1}\dot{e}'_{\rm q}+(1-\gamma_{\rm d1})\dot{\psi}''_{\rm d})}{(x''_{\rm d}x''_{\rm q}-R_{\rm s}^2)\imath^2} \\
        +&\frac{\bar{\imath}^{*}\bar{Z}_{\rm d}(-\gamma_{\rm q1}\dot{e}'_{\rm d}+(1-\gamma_{\rm q1})\dot{\psi}''_{\rm q})}{(x''_{\rm d}x''_{\rm q}-R_{\rm s}^2)\imath^2}
    \end{aligned}}
\end{equation}
The 6th-order model can be simplified by taking the assumptions $x''_{\rm d}=x'_{\rm d}$, $x''_{\rm q}=x'_{\rm q}$, and $T''_{\rm d0}=T''_{\rm q0}=0$, thus getting the so-called two-axis 4th-order model of the synchronous machine. The expressions in (\ref{eq:syn6axikappa}) simplify as:
\begin{equation}
  \begin{aligned}
  \label{eq:syn4xikappa}
  \bar{\xi}_{\rm a} &= \jmath\, \omega_{\rm r} + \frac{\bar{\imath}^{*}(\bar{Z}_{\rm d}(v_{\rm q}\omega_{\rm r}-\dot{e}'_{\rm d}) + \jmath \bar{Z}^{*}_{\rm q} (v_{\rm d}\omega_{\rm r}+\dot{e}'_{\rm q}))}{(x'_{\rm d}x'_{\rm q}-R_{\rm s}^2)\imath^2} \, , \\
  \bar{\kappa}_{\rho} &= \frac{\bar{\imath}^{*}(\bar{Z}_{\rm d}v_{\rm d}- \jmath \bar{Z}^{*}_{\rm q} v_{\rm q})}{(x'_{\rm d}x'_{\rm q}-R_{\rm s}^2)\imath^2} \, , \\
  \bar{\kappa}_{\omega} &= j\frac{\bar{\imath}^{*}(\jmath \bar{Z}_{\rm d} v_{\rm q} - \bar{Z}^{*}_{\rm q}v_{\rm d})}{(x'_{\rm d}x'_{\rm q}-R_{\rm s}^2)\imath^2} \, .
\end{aligned}
\end{equation}

The sought expression for $\bar{\chi}$ also simplifies as:
\begin{equation}\label{eq:syn4chi}
    \boxed{\begin{aligned}
        \ov{\chi}=\jmath\,(\omega_{\rm r}-\omega)&\left[\frac{\ov{\imath}^{*}(\ov{Z}_{\rm q}^{*}v_{\rm d}-\jmath\,\ov{Z}_{\rm d}v_{\rm q})}{(x'_{\rm d}x'_{\rm q}-R_{\rm s}^2)\imath^2}+1\right]\\
        -\rho&\left[\frac{\ov{\imath}^{*}(\jmath\,\ov{Z}_{\rm q}^{*}v_{\rm q}-\ov{Z}_{\rm d}v_{\rm d})}{(x'_{\rm d}x'_{\rm q}-R_{\rm s}^2)\imath^2}+1\right]\\
        +&\frac{\ov{\imath}^{*}(\jmath\,\ov{Z}_{\rm q}^{*}\dot{e}'_{\rm q}-\ov{Z}_{\rm d}\dot{e}'_{\rm d})}{(x'_{\rm d}x'_{\rm q}-R_{\rm s}^2)\imath^2}
    \end{aligned}}
\end{equation}

Further simplifying, for a 2nd-order classical model of the machine, namely
$R_{\rm s}=0$, $X'_{\rm q} = X'_{\rm d}$,
$T'_{\rm d0} = T'_{\rm q 0} = 0$, and $v_{\rm f}= \rm const.$, the
equation for $\bar{\chi}$ is:
\begin{equation}\label{eq:syn2chi}
    \boxed{\bar{\chi}=\left(\frac{-\jmath\,\bar{s}}{x'_{\rm d}\imath^2}+1\right)\left(-\rho +\jmath\,(\omega_{\rm r}-\omega)\right)}
\end{equation}
Imposing the ALS condition (\ref{eq:sync_ss}), we obtain:
\begin{equation}
\begin{aligned}
    \left(\frac{-\jmath\,\bar{s}}{X'_{\rm d}\imath^2}+1\right)\left(-\rho +\jmath\,(\omega_{\rm r}-\omega)\right)&\rightarrow 0 \, , \\
    \Rightarrow -\rho +\jmath\,(\omega_{\rm r}-\omega)&\rightarrow 0 \, , \\
    \Leftrightarrow \rho \rightarrow 0\, \land \, \omega&\rightarrow \omega_{\rm r} \, .
\end{aligned}
\end{equation}
Thus, a locally synchronized synchronous machine sets $\rho=0$ and $\omega$ equal to its internal state $\omega_{\rm r}$.

\subsection{ZIP loads}\label{subsubsec:zip}

The standard model of a ZIP load is:
\begin{align}
  p &=p_{\rm 0}\left(k_{\rm pp}+k_{\rm \imath p}v+k_{\rm zp}v^2\right)\, ,\\
  q &=q_{\rm 0}\left(k_{\rm pq}+k_{\rm \imath q}v+k_{\rm zq}v^2\right)\, ,
\end{align}
where the k-parameters represent the quota of the load that behaves as constant impedance, current or power.

For a constant impedance load, namely $k_{\rm zp}=k_{\rm zq}=1$, the results for $\bar{\xi}_{\rm a}$, $\bar{\kappa}_\rho$ and $\bar{\kappa}_\omega$ are:
\begin{equation}
  \bar{\xi}_{\rm a} = 0 \, , \qquad
  \bar{\kappa}_{\rho} = 1 \, , \qquad
  \bar{\kappa}_{\omega} = \jmath \, .
\end{equation}
Therefore:
\begin{equation}
    \boxed{\bar{\chi}=0}
\end{equation}
which indicates that the constant impedance load does not impose any constraint to $\rho$ or $\omega$ in order to achieve local synchronization.  This result also means that, by definition, it is always asymptotically locally synchronous according to the ALS condition (\ref{eq:sync_ss}).

For a constant current load ($k_{\rm \imath p}=k_{\rm \imath q}=1$):
\begin{equation}
  \bar{\xi}_{\rm a} = 0 \, , \qquad
  \bar{\kappa}_{\rho} =0 \, , \qquad
  \bar{\kappa}_{\omega} = \jmath \, .
\end{equation}
Therefore:
\begin{equation}
    \boxed{\bar{\chi}=-\rho} 
\end{equation}
Imposing the ALS condition (\ref{eq:sync_ss}), we obtain $\rho\rightarrow0$, which means that a locally synchronized constant current load sets $\rho=0$ while allowing $\omega$ to be free.

For a constant power load ($k_{\rm pp}=k_{\rm pq}=1$):
\begin{equation}
  \bar{\xi}_{\rm a} = 0 \, , \qquad
  \bar{\kappa}_{\rho} =-1 \, , \qquad
  \bar{\kappa}_{\omega} = \jmath \, .
\end{equation}
Therefore:
\begin{equation}
    \boxed{\bar{\chi}=-2\rho}
\end{equation}
Hence, the ALS condition for a constant power load requires $\rho =0$ but does not impose any constraint in $\omega$.

\subsection{Induction motors}

Consider the classical first-order model of an induction machine with the slip $\sigma$ as the state \cite{milanoscripting}:
\begin{equation}\label{eq:sigmadot}
    2H_{\rm m}\dot{\sigma}=\tau_{\rm m}-\tau_{\rm e} \, ,
\end{equation}
along with the algebraic equations:
\begin{equation}
\begin{aligned}
    \tau_{\rm e}&=\frac{\frac{r_{\rm R1}}{\sigma}v^2}{\left(r_{\rm S}+\frac{r_{\rm R1}}{\sigma}\right)^2+\left(x_{\rm S}+x_{\rm R1}\right)^2} \, ,\\
    p_{\rm e}&=\frac{\left(r_{\rm S}+\frac{r_{\rm R1}}{\sigma}\right)v^2}{\left(r_{\rm S}+\frac{r_{\rm R1}}{\sigma}\right)^2+\left(x_{\rm S}+x_{\rm R1}\right)^2} \, ,\\
    q_{\rm e}&=\frac{v^2}{x_\mu}+\frac{\left(x_{\rm S}+x_{\rm R1}\right)v^2}{\left(r_{\rm S}+\frac{r_{\rm R1}}{\sigma}\right)^2+\left(x_{\rm S}+x_{\rm R1}\right)^2} \, ,
\end{aligned}
\end{equation}
where $p_{\rm e}$, $q_{\rm e}$ are the active and reactive power drawn by the motor.
The CF of the current of the induction motor is:
\begin{equation}
    \bar{\xi}=\bar{\eta}-\frac{\dot{r}}{r}\left(\frac{r^2(x_{\rm t}^2-x^2)+\jmath\,rx_\mu(r^2-x^2-x_\mu x)}{z^2(r^2+x_{\rm t}^2)}\right) ,
\end{equation}
where we have used the notation:
\begin{equation}
\begin{aligned}
    r &= r_{\rm S}+\frac{r_{\rm R1}}{\sigma}, \quad
    x = x_{\rm S}+x_{\rm R1} \, ,\\
    x_{\rm t} &= x + x_\mu, \quad
    z^2 = r^2+x^2 \, ,
\end{aligned}
\end{equation}
and
\begin{equation}
    \dot{r}=-\frac{r_{\rm R1}}{\sigma^2}\dot{\sigma} \, ,
\end{equation}
where $\dot{\sigma}$ is known from (\ref{eq:sigmadot}).  Thus:
\begin{equation}
\begin{aligned}
    \bar{\xi}_{\rm a}&=-\frac{\dot{r}}{r}\left(\frac{r^2(x_{\rm t}^2-x^2)+\jmath\,rx_\mu(r^2-x^2-x_\mu x)}{z^2(r^2+x_{\rm t}^2)}\right) , \\
    \bar{\kappa}_\rho&= 1, \quad \bar{\kappa}_\omega=\jmath \, .
\end{aligned}
\end{equation}

The sought expression for $\bar{\chi}$ is:
\begin{equation}
    \boxed{\bar{\chi}=-\frac{\dot{r}}{r}\left(\frac{r^2(x_{\rm t}^2-x^2)+\jmath\,rx_\mu(r^2-x^2-x_\mu x)}{z^2(r^2+x_{\rm t}^2)}\right)}
\end{equation}

Imposing the ALS condition (\ref{eq:sync_ss}), we obtain:
\begin{equation}
\begin{aligned}
    -\frac{\dot{r}}{r}&\left(\frac{r^2(x_{\rm t}^2-x^2)+\jmath\,rx_\mu(r^2-x^2-x_\mu x)}{z^2(r^2+x_{\rm t}^2)}\right) \rightarrow 0 \, , \\
    &\Rightarrow \frac{\dot{r}}{r} \rightarrow 0 \, , \\
    &\Rightarrow \dot{\sigma} \rightarrow 0 \, .
\end{aligned}
\end{equation}
Therefore, an induction motor is locally synchronous as long as the torque balance is zero in (\ref{eq:sigmadot}).  This implies that the synchronization of the motor is solely determined by this internal balance, not by imposing any constraint on $\bar{\eta}$ at terminals. In fact, an induction motor operates without the need to rotate at the same speed as the frequency of the network. Interestingly, this aspect of its synchronization mechanism is similar to the one corresponding to a constant impedance load model. They both can be locally synchronized without linking $\bar{\eta}$ to an internal state. The difference lies in the internal balance requirement of the induction motor while the constant impedance load is always locally synchronized.

\subsection{Grid Following Inverter-Based Resources (GFL-IBR)}
\label{subsec:gflibr}

Consider a simplified GFL-IBR model with a synchronous reference Phase Locked Loop (PLL) whose frequency estimation is $\tilde{\omega}$, constant DC voltage $v_{\rm dc0}$, and PI controllers for the internal current control loops ($K_{\rm p}$, $K_{\rm i}$) with fixed references $\bar{\imath}_{\rm ref}=\imath_{\rm dref}+\jmath\,\imath_{\rm qref}$.  The modulated signal is $\bar{m}$, and $\bar{z}_{\rm f}$, $\bar{y}_{\rm f}$ are the output filter series impedance and shunt conductance, respectively.  The set of differential equations is as follows:
\begin{equation}
  \begin{aligned}
    \dot{x}_{\rm d}&=K_{\rm i}(\imath_{\rm dref}-\imath_{\rm dm})\,,\\
    \dot{x}_{\rm q}&=K_{\rm i}(\imath_{\rm qref}-\imath_{\rm qm})\,,\\
    T_{\rm m }\dot{\imath}_{\rm dm}&=\imath_{\rm d}-\imath_{\rm dm}\,,\\
    T_{\rm m }\dot{\imath}_{\rm qm}&=\imath_{\rm q}-\imath_{\rm qm}\,,\\
    \dot{x}_{\rm pll}&=K_{\rm i_{\rm pll}}v_{\rm q}\,,\\
    \dot{\tilde{\theta}}&=\Delta\omega_{\rm pll}\,,
  \end{aligned}
\end{equation}
and the set of algebraic equations is:
\begin{equation}
  \begin{aligned}
    \bar{m}&=\bar{x}+K_{\rm p}(\bar{\imath}_{\rm ref}-\bar{\imath}_{\rm m})\,,\\
    \bar{v}&=\bar{m}v_{\rm dc0}-\bar{z}_{\rm f}(\bar{\imath}+\bar{y}_{\rm f}\bar{v})\,,\\
    \Delta\omega_{\rm pll}&=K_{\rm p_{\rm pll}}v_{\rm q}+x_{\rm pll}\, ,\\
    \tilde{\omega}&=\Delta\omega_{\rm pll}+\omega_{\rm ref}\, ,\\
    \bar{v} &= v_{\rm d} + \jmath \, v_{\rm q} = v\cos(\theta-\tilde{\theta}) +
    \jmath\, v\sin(\theta-\tilde{\theta}) \, ,\\
    \bar{\imath} &= \imath_{\rm d} + \jmath \, \imath_{\rm q} =
    \imath\cos(\theta-\tilde{\theta}) + \jmath\,\imath\sin(\theta-\tilde{\theta})
    \,.
  \end{aligned}
\end{equation}
The results for $\bar{\xi}_{\rm a}$, $\bar{\kappa}_\rho$ and $\bar{\kappa}_\omega$ are:
\begin{equation}
  \begin{aligned}
    \label{eq:gflxikappa}
    \bar{\xi}_{\rm a} &=  \frac{\bar{m}v_{\rm dc0}}{\bar{z}_{\rm f}\bar{\imath}}\left(\frac{\dot{m}}{m}+\jmath\, (\dot{\alpha}+\tilde{\omega})\right) , \\
    \bar{\kappa}_{\rho} &=  1 - \frac{\bar{m}v_{\rm dc0}}{\bar{z}_{\rm f}\bar{\imath}}\, , \quad
    \bar{\kappa}_{\omega} = \jmath\,\left(1 - \frac{\bar{m}v_{\rm dc0}}{\bar{z}_{\rm f}\bar{\imath}}\right) .
  \end{aligned}
\end{equation}
The sought expression for $\bar{\chi}$ is:
\begin{equation}
    \boxed{
    \bar{\chi}=\frac{\bar{m}\, v_{\rm dc0}}{\bar{z_{\rm f}}\, \bar{\imath}}\left(\frac{\dot{m}}{m}-\rho+\jmath\,(\dot{\alpha}+\tilde{\omega}-\omega)\right)}
\end{equation}
where:
\begin{equation}
  \begin{aligned}
    \frac{\dot{m}}{m}&=\frac{m_{\rm d}\dot{m}_{\rm d}+m_{\rm q}\dot{m}_{\rm q}}{m_{\rm d}^2+m_{\rm q}^2} \, , \\
    \dot{\alpha}&=\frac{m_{\rm d}\dot{m}_{\rm q}-m_{\rm q}\dot{m}_{\rm d}}{m_{\rm d}^2+m_{\rm q}^2} \, ,
  \end{aligned}
\end{equation}
and
\begin{equation}
\begin{aligned}
    \dot{m}_{\rm d}&=K_{\rm i}(\imath_{\rm dref}-\imath_{\rm d})-\frac{K_{\rm p}}{T_{\rm m}}(\imath_{\rm d}-\imath_{\rm dm}) \, , \\
    \dot{m}_{\rm q}&=K_{\rm i}(\imath_{\rm qref}-\imath_{\rm q})-\frac{K_{\rm p}}{T_{\rm m}}(\imath_{\rm q}-\imath_{\rm qm}) \, ,
\end{aligned}
\end{equation}
Applying the ALS condition (\ref{eq:sync_ss}), we obtain:
\begin{equation}
\begin{aligned}
    \frac{\bar{m}\, v_{\rm dc0}}{\bar{z_{\rm f}}\, \bar{\imath}}\left(\frac{\dot{m}}{m}-\rho+\jmath\,(\tilde{\omega}+\dot{\alpha}-\omega)\right)&\rightarrow 0 \, , \\
    \Rightarrow \frac{\dot{m}}{m}-\rho+\jmath\,(\tilde{\omega}+\dot{\alpha}-\omega)&\rightarrow 0 \, , \\
    \Leftrightarrow \rho\rightarrow \frac{\dot{m}}{m}\, \land \, \omega&\rightarrow \tilde{\omega}+\dot{\alpha} \, .
\end{aligned}
\end{equation}
Therefore, a locally synchronized GFL-IBR drives $\rho$ to $\frac{\dot{m}}{m}$ and $\omega$ to $(\tilde{\omega}+\dot{\alpha})$.

\subsection{Grid Forming Inverter-Based Resources (GFM-IBR)}

Consider a GFM-IBR according to the WECC REGFM\_A1 model \cite{WECCGFM}:
\begin{equation}
  \begin{aligned}
    \dot{e}&=K_{\rm i}(v_{\rm ref}-v_{\rm m})-\frac{K_{\rm p}}{T_{\rm v}}(v_{\rm m}-v) \, , \\
    \dot{\delta}&=\Omega_b(\omega_{\rm gfm}-1)\, ,
  \end{aligned}
\end{equation}
along with the algebraic equations:
\begin{equation}
  \begin{aligned}
    \omega_{\rm gfm}&=m_{\rm p}(p_{\rm ref}-p_{\rm m})+1\, , \\
    \bar{v}&=\bar{e}-\bar{z}_{\rm t}\bar{\imath}\, .
  \end{aligned}
\end{equation}
The results for $\bar{\xi}_{\rm a}$, $\bar{\kappa}_\rho$ and $\bar{\kappa}_\omega$ are:
\begin{equation}
  \begin{aligned}
    \label{eq:gfmxikappa}
    \bar{\xi}_{\rm a} &=  \frac{\bar{e}}{\bar{z}_{\rm t}\bar{\imath}}\left(\frac{\dot{e}}{e}+\jmath\, \omega_{\rm gfm}\right) , \\
    \bar{\kappa}_{\rho} &=  1 - \frac{\bar{e}}{\bar{z}_{\rm t}\bar{\imath}}\, , \quad
    \bar{\kappa}_{\omega} = \jmath\,\left(1 - \frac{\bar{e}}{\bar{z}_{\rm t}\bar{\imath}}\right)  .
  \end{aligned}
\end{equation}
The sought expression for $\bar{\chi}$ is:
\begin{equation}
  \boxed{
    \bar{\chi} = \frac{\bar{e}}{\bar{z}_{\rm t}\,
      \bar{\imath}}\left(\frac{\dot{e}}{e}-\rho+\jmath\,
    (\omega_{\rm gfm}-\omega)\right)}
\end{equation}
Imposing the ALS condition (\ref{eq:sync_ss}), we obtain:
\begin{equation}
  \begin{aligned}
    \frac{\bar{e}}{\bar{z}_{\rm t}\, \bar{\imath}}\left(\frac{\dot{e}}{e}-\rho+\jmath\,(\omega_{\rm gfm}-\omega)\right)&\rightarrow 0 \, ,\\
    \Rightarrow \frac{\dot{e}}{e}-\rho+\jmath\,(\omega_{\rm gfm}-\omega)&\rightarrow 0 \, , \\
    \Leftrightarrow \rho\rightarrow \frac{\dot{e}}{e}\, \land \, \omega&\rightarrow \omega_{\rm gfm} \, .
  \end{aligned}
\end{equation}
Therefore, a locally synchronized GFM-IBR drives $\rho$ to $\frac{\dot{e}}{e}$ and $\omega$ to $\omega_{\rm gfm}$.

\section{Case Studies}
\label{sec:examples}

This section discusses the proposed concept of local synchronization in different scenarios using benchmark systems composed of common power system devices.  Simulations were carried out using the Python-based software tool Dome \cite{dome}.

\subsection{Circuit-based Illustrative Example}
\label{subsec:emt}

\begin{figure}[hbtp]
    \centering
    \includegraphics[width = 0.45\textwidth]{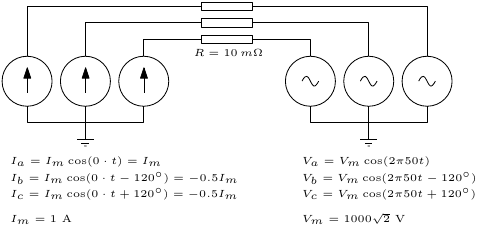}
    \caption{Circuit diagram. Example \ref{subsec:emt}.}
    \label{fig:emt_circuit}
\end{figure}

\begin{figure}[hbtp]
    \centering
    \includegraphics[width = 0.45\textwidth]{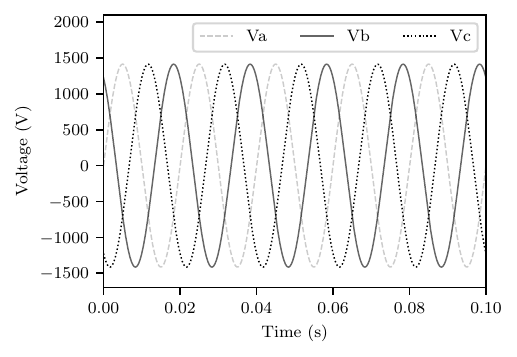}
    \vspace{-4mm}
    \caption{Three-phase voltage trajectories. Example \ref{subsec:emt}.}
    \label{fig:vabc_emt}
    \vspace{-2mm}
\end{figure}

\begin{figure}[hbtp]
    \centering
    \includegraphics[width = 0.45\textwidth]{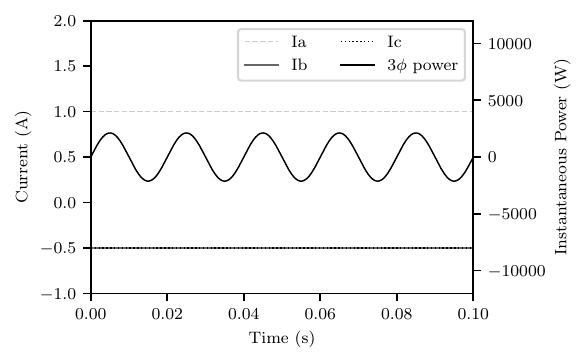}
    \vspace{-4mm}
    \caption{Current and power trajectories. Example \ref{subsec:emt}.}
    \label{fig:iabcp_emt}
    \vspace{-2mm}
\end{figure}

\begin{figure}[hbtp]
    \centering
    \includegraphics[width = 0.45\textwidth]{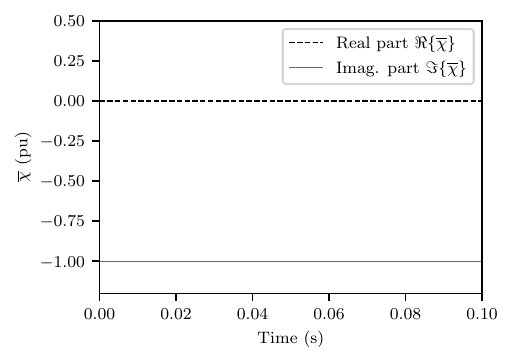}
    \vspace{-4mm}
    \caption{Admittance CF $\bar{\chi}$. Example \ref{subsec:emt}.}
    \label{fig:chi_emt}
    \vspace{-2mm}
\end{figure}

We first illustrate the independence of local synchronization from stability through a simple illustrative example.  Consider the circuit shown in Fig.~\ref{fig:emt_circuit}.  The voltage source is ideally balanced and at nominal frequency, and the current source injects a constant (DC) current.  Fig. \ref{fig:vabc_emt} shows the three-phase voltage at the terminal node of the current source, whereas Fig. \ref{fig:iabcp_emt} shows the constant three-phase current on the left-hand side axis, and the instantaneous power on the right-hand side axis.  Note that the device is not able to transmit active power on average as the voltage and current do not have the same frequency.  The trajectories of $\bar{\chi}$ for the current source are shown in Fig. \ref{fig:chi_emt}.  This circuit is stable, but the device not locally synchronous according to our definition.  


\color{black}

\subsection{Synchronous machine infinite bus (SMIB) system}
\label{subsec:smib}

\begin{figure}[hbtp]
    \centering
    \includegraphics[scale=0.9]{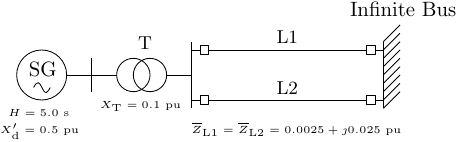}
    \caption{Single line diagram of the SMIB system.}
    \label{fig:smib}
\end{figure}

Consider the classical situation of a synchronous machine connected to an equivalent network. The machine is represented using the classical model and is connected to an infinite bus through a transformer and two parallel lines.  The single-line diagram of the system is shown in Fig.~\ref{fig:smib}.

A three-phase short circuit is applied at simulation time $t=1$ s in the middle of L2 and cleared at $t_{\rm cl}$ by opening the breakers of the line.

\begin{figure}[hbtp]
    \centering
    \includegraphics[width = 0.45\textwidth]{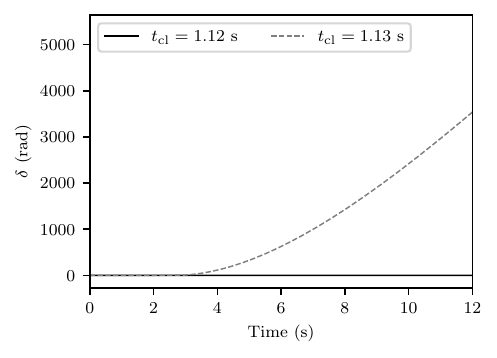}
    \vspace{-4mm}
    \caption{Internal angle of the SG.  Example \ref{subsec:smib}.}
    \label{fig:delta_smib}
\end{figure}

Figure \ref{fig:delta_smib} shows the response of the internal angle of the machine in two cases: $t_{\rm cl}=1.12$ s and $t_{\rm cl}=1.13$ s. The critical clearing time is thus $\sim 130$ ms.

\begin{figure}[hbtp]
    \centering
    \includegraphics[width = 0.45\textwidth]{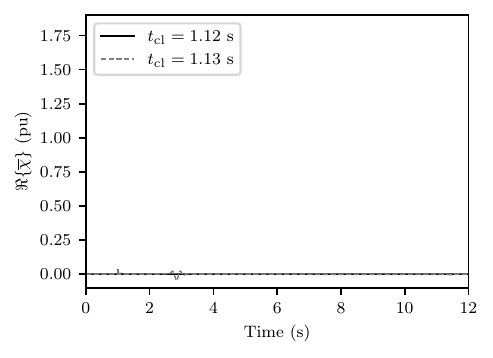}
    \vspace{-4mm}
    \caption{Real part of $\bar{\chi}$. Example \ref{subsec:smib}.}
    \label{fig:rho_smib}
    \vspace{-2mm}
\end{figure}

\begin{figure}[hbtp]
    \centering
    \includegraphics[width = 0.45\textwidth]{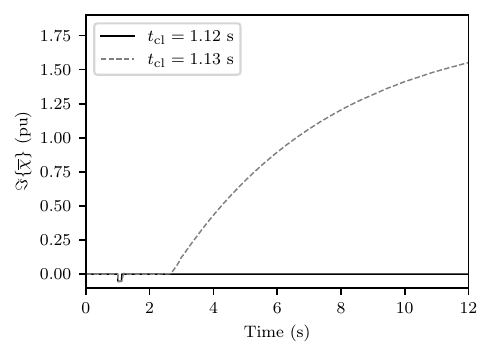}
    \vspace{-4mm}
    \caption{Imaginary part of $\bar{\chi}$. Example \ref{subsec:smib}.}
    \label{fig:om_smib}
    \vspace{-2mm}
\end{figure}

The trajectories of the real and imaginary parts of $\bar{\chi}$ are shown in Fig. \ref{fig:rho_smib} and Fig. \ref{fig:om_smib}, respectively. The machine achieves asymptotic local synchronization when $t_{\rm cl}=1.12$ s as $\bar{\chi}$ converges to 0. In turn, it loses local synchronism when $t_{\rm cl}=1.13$ s as $\bar{\chi}$ diverges, particularly the imaginary part of it because the problem is associated with an angle drift. The results are under expectations in the case of the well-studied phenomena of loss of synchronism of a synchronous machine.  In this case, the conditions for achieving local synchronization and stability are intrinsically intertwined.  That is, every time the machine loses local synchronization the system is unstable, and every time the machine achieves local synchronization, the system is stable.

\subsection{Kundur's two-areas system}\label{subsec:kundur}

\begin{figure}[hbtp]
    \centering
    \includegraphics[scale=0.9]{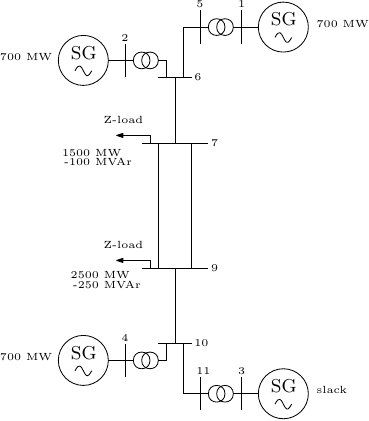}
    \caption{Single line diagram of the Kundur's two-areas system.}
    \label{fig:kundur}
    \vspace{-2mm}
\end{figure}

Consider the two-area system proposed in \cite{kundur}.  This system is composed of four synchronous machines, two in each area, which are modeled using the 4th-order model described in Section \ref{subsubsec:syn}. The loads are modeled as constant impedances. The single-line diagram of the system is shown in Fig.~\ref{fig:kundur}.

A three-phase short circuit is applied in the middle of one of the circuits of line 07-09. The fault is applied at simulation time $t=1$ s and cleared after 120 ms by opening the faulted circuit.

\begin{figure}[hbtp]
    \centering
    \includegraphics[width = 0.45\textwidth]{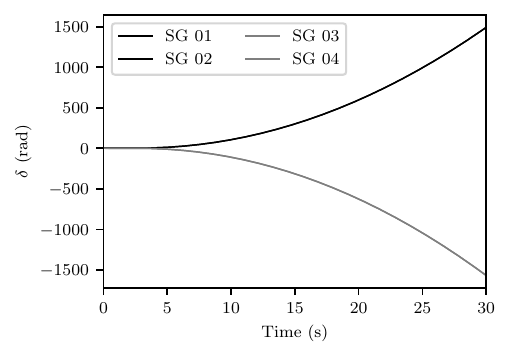}
    \vspace{-4mm}
    \caption{Internal angles of the SGs.  Example \ref{subsec:kundur}.}
    \label{fig:delta_kundur}
    \vspace{-2mm}
\end{figure}

\begin{figure}[hbtp]
    \centering
    \includegraphics[width = 0.45\textwidth]{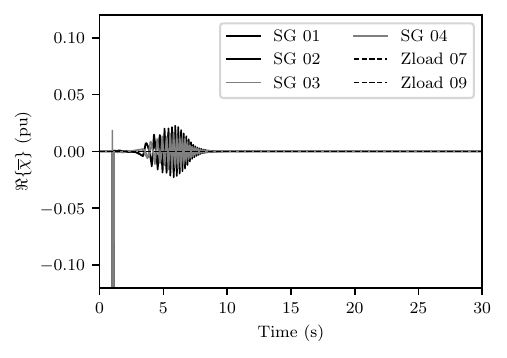}
    \vspace{-4mm}
    \caption{Real part of $\bar{\chi}$.  Example \ref{subsec:kundur}.}
    \label{fig:rho_kundur}
    \vspace{-2mm}
\end{figure}

\begin{figure}[hbtp]
    \centering
    \includegraphics[width = 0.45\textwidth]{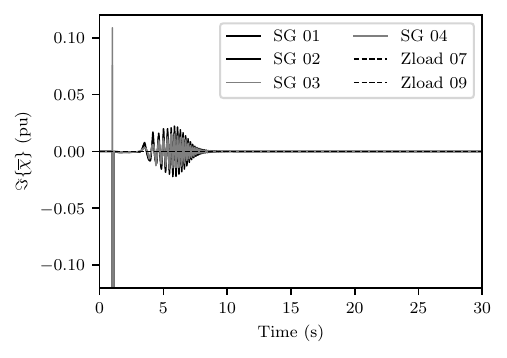}
    \vspace{-4mm}
    \caption{Imaginary part of $\bar{\chi}$. Example \ref{subsec:kundur}.}
    \label{fig:om_kundur}
    \vspace{-2mm}
\end{figure}

The contingency forces the two groups of machines to drift away, separating the system and leading to an unstable response. The time-domain response of the angles of the generators is shown in Fig.~\ref{fig:delta_kundur}. Also, the trajectories of the real and imaginary part of $\bar{\chi}$ for all the devices are shown in Fig. \ref{fig:rho_kundur} and Fig.~\ref{fig:om_kundur}, respectively.

Note that even though the system's response is unstable and there is an angular separation between the two areas, the results evidence that the generators still achieve asymptotic local synchronization as $\bar{\chi}$ tends to 0. This is because the root of the problem is on the transmission side, triggered by the inter-area mode, not in any of the machines particularly. 

\begin{figure}[hbtp]
  \centering
  \includegraphics[width = 0.45\textwidth]{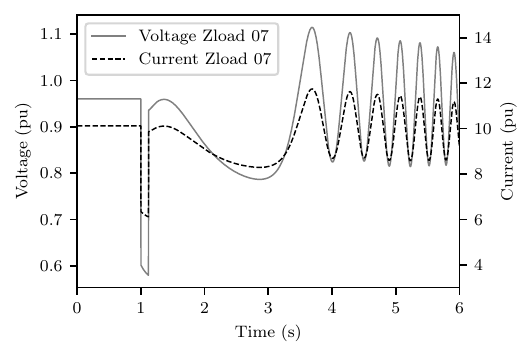}
  \vspace{-4mm}
  \caption{Voltage and current magnitude of Zload 07.  Example \ref{subsec:kundur}.}
  \label{fig:iv_kundur}
  \vspace{-2mm}    
\end{figure}

For constant impedance loads, $\bar{\chi}$ remains perfectly constant, which implies that they permanently keep local synchronization no matter the system response.  This result is consistent with their entirely passive and always-synchronized nature.  Figure \ref{fig:iv_kundur} shows the voltage and current magnitude of load ``Zload 07,'' in the first six seconds of simulation.  The comparison of Figs.~\ref{fig:rho_kundur} and \ref{fig:iv_kundur} highlights the convenience of observing $\bar{\chi}$ as a metric of local synchronization.

\subsection{Induction machine to infinite bus system}
\label{subsec:ind1}

\begin{figure}[hbtp]
    \centering
    \includegraphics[scale=0.9]{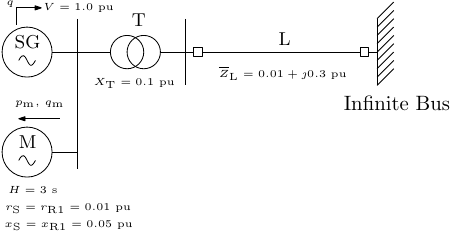}
    \caption{Single line diagram of the induction machine system.}
    \label{fig:ind1}
\end{figure}

Consider a system composed of a first-order classical model of an induction motor connected to an infinite bus through a transformer and a line. There is also a synchronous condenser regulating the terminal voltage of the motor at 1.0 pu. The single-line diagram of the system is shown in Fig.~\ref{fig:ind1}.

\begin{figure}[hbtp]
    \centering
    \includegraphics[width = 0.45\textwidth]{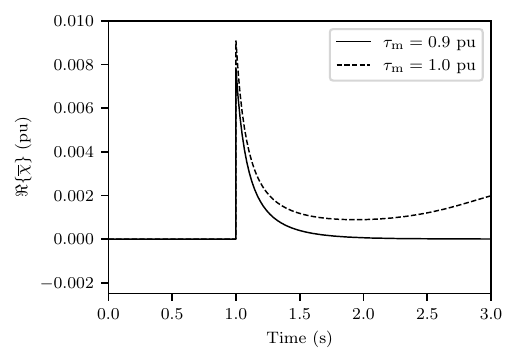}
    \vspace{-4mm}
    \caption{Real part of $\bar{\chi}$.  Example \ref{subsec:ind1}.}
    \label{fig:rho_ind1}
    \vspace{-2mm}
\end{figure}

\begin{figure}[hbtp]
    \centering
    \includegraphics[width = 0.45\textwidth]{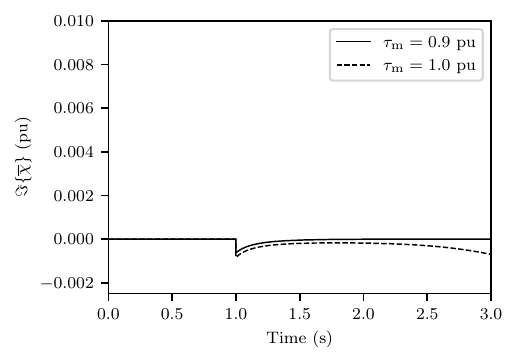}
    \vspace{-4mm}
    \caption{Imaginary part of $\bar{\chi}$.  Example \ref{subsec:ind1}.}
    \label{fig:om_ind1}
    \vspace{-2mm}
\end{figure}

The synchronous condenser is suddenly disconnected at simulation time $t=1$ s. 
The trajectories of the real and imaginary parts of the induction motor $\bar{\chi}$ are shown in Figs.~\ref{fig:rho_ind1} and \ref{fig:om_ind1}, respectively.  The motor mechanical load is modeled as constant torque, and two cases are considered.  With $\tau_{\rm m}=0.9$ pu, the motor rides through the sudden disconnection of the condenser without losing synchronism. In contrast, with a higher load of $\tau_{\rm m}=1.0$ pu, the system response is not stable and the motor stalls, also losing synchronism as $\bar{\chi}$ diverges.  The phenomenon is mostly reflected in the real part of $\bar{\chi}$ as the nature of the problem is associated with the inability of the system to supply the motor reactive power demand, leading to a voltage collapse at its terminals.

\subsection{IEEE 14-bus system}
\label{subsec:ieee14}

\begin{figure}[hbtp]
  \centering
  \includegraphics[width=0.7\linewidth]{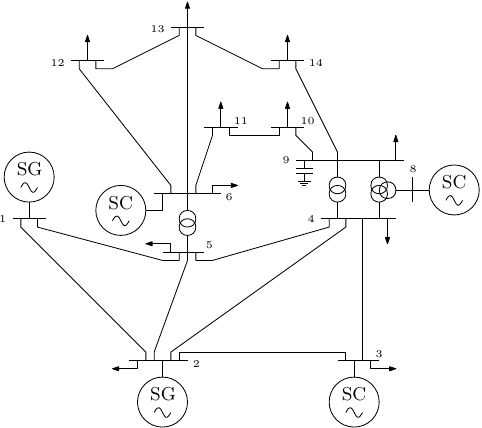}
  \caption{Single line diagram of the IEEE 14-bus system.}
  \label{fig:ieee14}
\end{figure}

Consider the well-known IEEE 14-bus system whose single-line diagram is shown in Fig.~\ref{fig:ieee14}.  The synchronous machines are represented using a 6th-order model with IEEE-type DC1 voltage regulators and no power system stabilizers.  Also, the two generators are equipped with turbine governors. Details of the system's data and base operating conditions can be found in \cite{milanoscripting}.

This benchmark system has been widely used to study Hopf bifurcations and limit cycles (see, for example, \cite{bizarri} and the references therein).  In particular, a Hopf bifurcation and following limit cycle occur when the system operates under overloading conditions.  We consider an operating scenario with an overloading level of 20\% relative to the base operating conditions. 

A three-phase short circuit is applied at bus 14 at simulation time $t=1$ s and cleared at $t_{\rm cl}$.  Two different cases in terms of the clearing time are considered, $60$ ms and $120$ ms.  Following the fault, the system trajectory ends up on a limit cycle.  Although the limit cycle is mathematically stable as it can be concluded by calculating the Floquet multipliers of the monodromy matrix of the trajectory of the system \cite{bizarri}, such stationary operating conditions are not acceptable in practice.

\begin{figure}[hbtp]
  \centering
  \includegraphics[width = 0.45\textwidth]{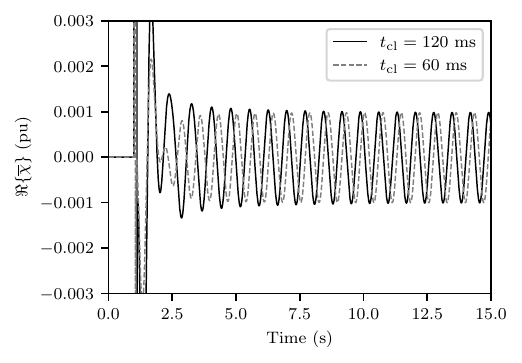}
  \vspace{-4mm}
  \caption{Real part of $\bar{\chi}$.  Example \ref{subsec:ieee14}.}
  \label{fig:rho_ieee14}
  \vspace{-2mm}
\end{figure}

\begin{figure}[hbtp]
  \centering
  \includegraphics[width = 0.45\textwidth]{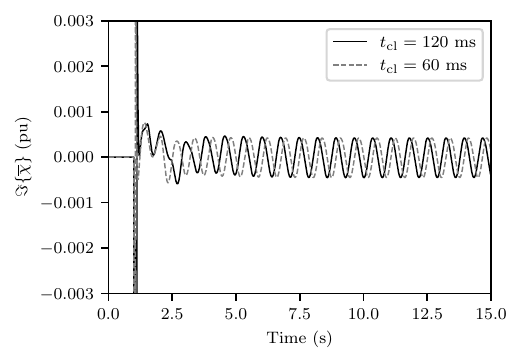}
  \vspace{-4mm}
  \caption{Imaginary part of $\bar{\chi}$.  Example \ref{subsec:ieee14}.}
  \label{fig:om_ieee14}
  \vspace{-2mm}
\end{figure}

The SG connected at bus 1 is taken as an example to evaluate local synchronization. The trajectories of the real and imaginary parts of $\bar{\chi}$ are shown in Figs.~\ref{fig:rho_ieee14} and \ref{fig:om_ieee14}, respectively.  Note that, while the system trajectories do not diverge, the generator does not achieve either ALS or BLS.  Not even the less restrictive BLS condition (\ref{eq:sync_eps}) is satisfied in fact, as there is no $\delta$ for an arbitrarily small $\epsilon$ (e.g., $\epsilon=1^{-4}$).  The trajectories of $\bar{\chi}$ tend to oscillate with a minimum magnitude given by the limit cycle, thus not allowing the devices to achieve local synchronization despite the system being mathematically stable.

\subsection{GFL-IBR and series line compensation system}
\label{subsec:pllc}

\begin{figure}
  \centering
  \includegraphics[scale=0.9]{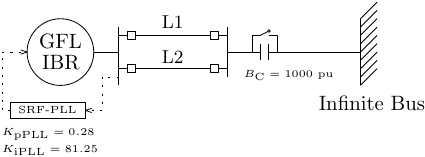}
  \caption{Single line diagram of the GFL-IBR and series line compensation system.}
  \label{fig:pllc}
\end{figure}

The last simulation example is a typical scenario of resonance between the series compensation of a weak line and a poorly tuned SRF-PLL of a standard grid-following controlled converter.  The single-line diagram of the system is shown in Fig. \ref{fig:pllc}.  The model used to represent the converter is the one described in Section \ref{subsec:gflibr}. 

\begin{figure}
    \centering
    \includegraphics[width=0.9\linewidth]{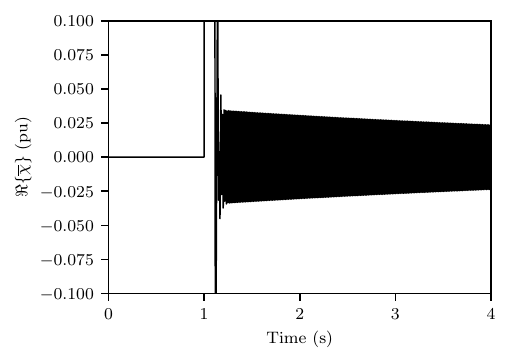}
    \vspace{-4mm}
    \caption{Real part of $\bar{\chi}$. Example \ref{subsec:gflibr}.}
    \label{fig:rho_pllc}
    \vspace{-2mm}
\end{figure}

\begin{figure}
    \centering
    \includegraphics[width=0.9\linewidth]{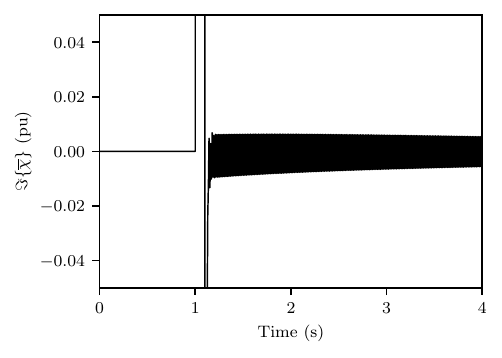}
    \vspace{-4mm}
    \caption{Imaginary part of $\bar{\chi}$. Example \ref{subsec:gflibr}.}
    \label{fig:om_pllc}
    \vspace{-2mm}
\end{figure}

A three-phase fault at L2 close to the terminals of the converter is applied at simulation time $t=1$ s and cleared after 100 ms by opening the line.  Local synchronization of the converter is evaluated by observing the trajectories of $\bar{\chi}$.  The real and imaginary parts are shown in Figs.~\ref{fig:rho_pllc} and \ref{fig:om_pllc}, respectively.

The dynamic response is characterized by poorly damped high-frequency oscillations caused by the interaction between the series capacitor and the PLL.  The trajectories of $\bar{\chi}$ consistently capture the phenomenon, evidencing that the converter is not entirely coherent with the rest of the system, as the oscillations decay slowly.  Nevertheless, it still achieves asymptotic local synchronization as $\bar{\chi}$ tends to zero. 

\section{Conclusions}
\label{sec:conclusion}

The paper proposes a novel concept of local synchronization based on the difference between the complex frequency of the voltage and current injected at device terminals. It is formalized through two definitions that account for bounded and asymptotic local synchronization.

A taxonomy of the synchronization mechanisms of common power system devices is provided.  This description is done systematically and leads to analytical explicit expressions.  For instance, we show that synchronous machines, grid-following and grid-forming models of converters require the CF of the voltage at terminals to be equal to certain internal states to synchronize, e.g., $\omega$ has to match the rotor speed of synchronous machines.  On the other hand, constant shunt impedances and induction machines do not constrain the CF of the grid voltage.  

The proposed definition of local synchronization is fully independent from and complements stability.  That is, for a proper power system operation, a system has to be stable, and every device has to be local synchronous.  In some cases, the manifestation of local loss of synchronism may reveal the origin of system motion that is not acceptable in practice.  However, a device not achieving local synchronization does not imply that the cause is local.  In fact, system-wide dynamic interactions may trigger loss of stability but not of local synchronism, see case \ref{subsec:kundur}; or local losses of synchronism but not of stability, see case \ref{subsec:ieee14}. 

The structural differences between the two aforementioned cases suggest that the proposed definition can help revisit the definitions of stability of power systems.  This is one of the research directions that we will pursue in future work. The wide variety of shapes observed in the trajectories of $\bar{\chi}$ also motivates us to further characterize local synchronization by exploring the information the transient behavior of this key complex variable can reveal.  Future work will focus, in particular, on estimating $\bar{\chi}$ based on the measurements of voltages and currents and exploit this quantity for control purposes.  We are also exploring applications of the analytical outcomes of the paper to improve the controllers of the devices discussed in this work, in particular, GFL- and GFM-IBRs.  Finally, another future research line is to extend the application of local synchronization from single devices to network areas.  This will require exploring how our framework can be adapted to study the synchronization mechanism of a portion of the grid composed by heterogeneous devices in a general and systematic way.

\newpage

\begin{IEEEbiography}[{\includegraphics[width=1in, height=1.25in,
    clip, keepaspectratio]{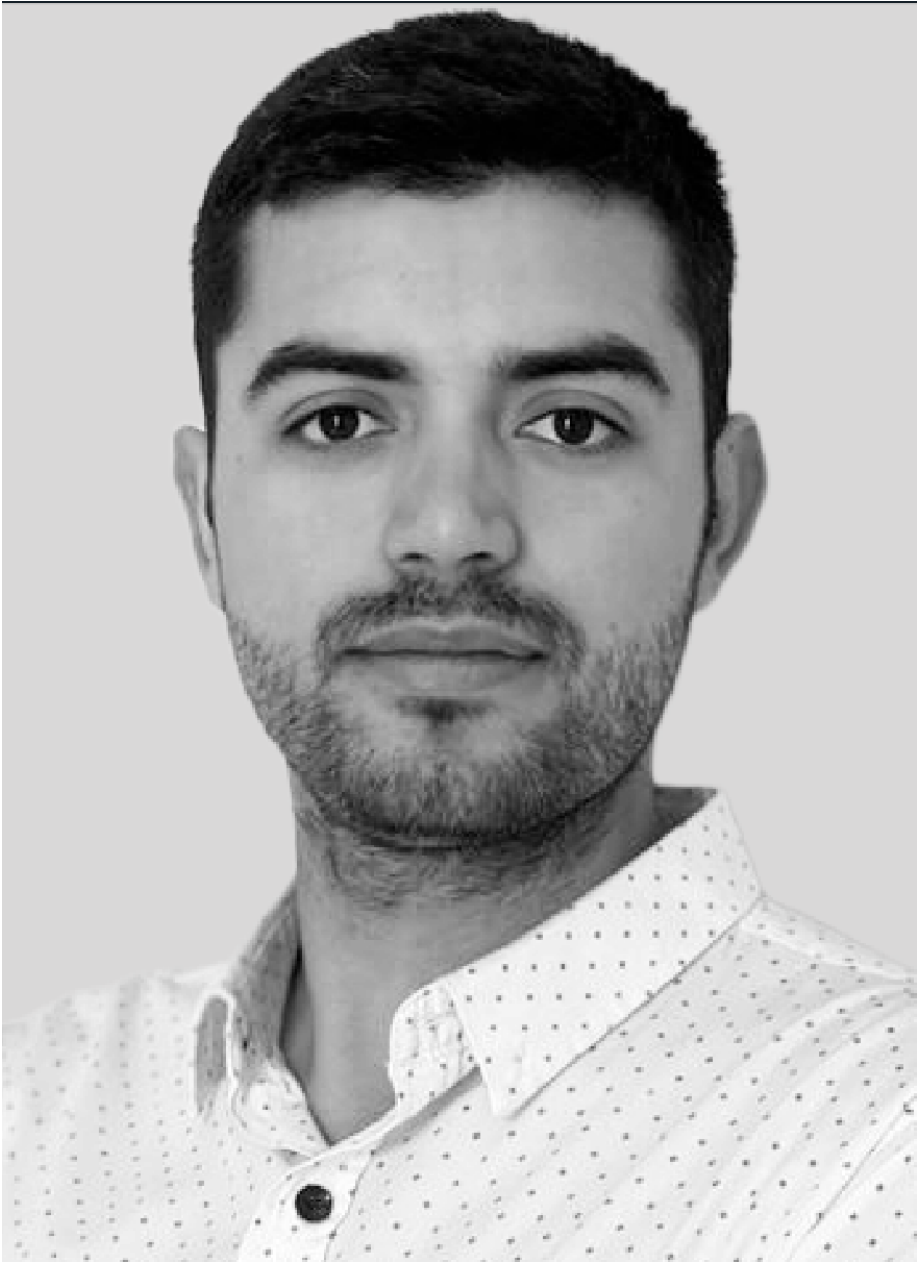}}] {Ignacio Ponce} received from the University of Chile the BSc.~and MSc.~degree in Electrical Engineering in 2019 and 2022, respectively.  He is currently pursuing a Ph.D in Electrical Engineering at University College Dublin, Ireland.  His research interests include power system modeling, control and stability analysis.
\end{IEEEbiography}

\begin{IEEEbiography}[{\includegraphics[width=1in, height=1.25in,
    clip, keepaspectratio]{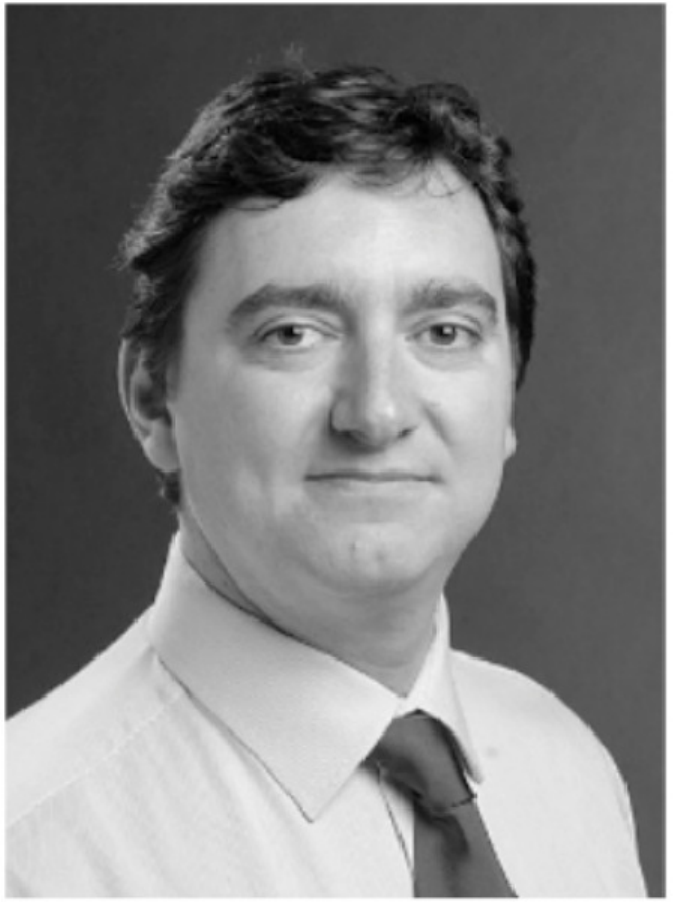}}] {Federico Milano} (F'16) received from the Univ.~of Genoa, Italy, the Ph.D.~in Electrical Engineering 2003.  In 2013, he joined the University College Dublin, Ireland, where he is currently a full professor.  He is Chair of the IEEE Power System Stability Controls Subcommittee, IET Fellow, IEEE PES Distinguished Lecturer, Chair of the Technical Program Committee of the PSCC 2024, Senior Editor of the IEEE Transactions on Power Systems, Member of the Cigr{\'e} Irish National Committee, and Co-Editor in Chief of the IET Generation, Transmission \& Distribution.  His research interests include power system modeling, control and stability analysis.
\end{IEEEbiography}

\vfill


\begin{thebibliography}{10}

\bibitem{sauerpai}
P.~Sauer and M.~Pai, {\em Power System Dynamics and Stability}.
\newblock Prentice Hall, 1998.

\bibitem{taskforcestab}
N.~Hatziargyriou {\em et~al.}, ``Definition and classification of power system
  stability – revisited \& extended,'' {\em IEEE Transactions on Power
  Systems}, vol.~36, no.~4, pp.~3271--3281, 2021.

\bibitem{paganini}
F.~Paganini and E.~Mallada, ``Global analysis of synchronization performance
  for power systems: Bridging the theory-practice gap,'' {\em IEEE Transactions
  on Automatic Control}, vol.~65, no.~7, pp.~3007--3022, 2020.

\bibitem{colombino}
M.~Colombino, D.~Groß, J.-S. Brouillon, and F.~Dörfler, ``Global phase and
  magnitude synchronization of coupled oscillators with application to the
  control of grid-forming power inverters,'' {\em IEEE Transactions on
  Automatic Control}, vol.~64, no.~11, pp.~4496--4511, 2019.

\bibitem{yangspectral}
P.~Yang, F.~Liu, Z.~Wang, S.~Wu, and H.~Mao, ``Spectral analysis of network
  coupling on power system synchronization with varying phases and voltages,''
  in {\em 2020 Chinese Control And Decision Conference (CCDC)}, pp.~880--885,
  2020.

\bibitem{synchronizability}
C.~Yu and W.~Xiao, ``Synchronizability and synchronization rate of ac
  microgrids: A topological perspective,'' {\em IEEE Transactions on Network
  Science and Engineering}, vol.~11, no.~2, pp.~1424--1441, 2024.

\bibitem{cfsyncdorfler}
X.~He, V.~H{\"a}berle, and F.~D{\"o}rfler, ``Complex-frequency synchronization
  of converter-based power systems,'' 2024.

\bibitem{augmented}
P.~Yang, F.~Liu, T.~Liu, and D.~J. Hill, ``Augmented synchronization of power
  systems,'' {\em IEEE Transactions on Automatic Control}, vol.~69, no.~6,
  pp.~3673--3688, 2024.

\bibitem{ComplexFreq}
F.~Milano, ``Complex frequency,'' {\em IEEE Transactions on Power Systems},
  vol.~37, no.~2, pp.~1230--1240, 2022.

\bibitem{kuramoto}
Y.~Kuramoto, ``Self-entrainment of a population of coupled non-linear
  oscillators,'' in {\em International Symposium on Mathematical Problems in
  Theoretical Physics} (H.~Araki, ed.), (Berlin, Heidelberg), pp.~420--422,
  Springer Berlin Heidelberg, 1975.

\bibitem{dVOC}
M.~Colombino, D.~Groß, and F.~Dörfler, ``Global phase and voltage
  synchronization for power inverters: A decentralized consensus-inspired
  approach,'' in {\em 2017 IEEE 56th Annual Conference on Decision and Control
  (CDC)}, pp.~5690--5695, 2017.

\bibitem{taxonomy}
D.~Moutevelis, J.~Roldán-Pérez, M.~Prodanovic, and F.~Milano, ``Taxonomy of
  power converter control schemes based on the complex frequency concept,''
  {\em IEEE Transactions on Power Systems}, vol.~39, no.~1, pp.~1996--2009,
  2024.

\bibitem{modalprop}
D.~Moutevelis, G.~Tzounas, J.~Roldán-Pérez, and F.~Milano, ``Modal
  propagation analysis with participation factors of complex frequency
  variables,'' {\em Electric Power Systems Research}, vol.~230, p.~110295,
  2024.

\bibitem{cfdf}
I.~Ponce and F.~Milano, ``Complex frequency divider,'' {\em Electric Power
  Systems Research}, vol.~234, p.~110662, 2024.

\bibitem{hybridcf}
I.~Ponce and F.~Milano, ``Modeling hybrid ac/dc power systems with the complex
  frequency concept,'' {\em IEEE Transactions on Power Systems}, vol.~39,
  no.~4, pp.~6004--6013, 2024.

\bibitem{enhancing}
F.~Milano, B.~Alhanjari, and G.~Tzounas, ``Enhancing frequency control through
  rate of change of voltage feedback,'' {\em IEEE Transactions on Power
  Systems}, vol.~39, no.~1, pp.~2385--2388, 2024.

\bibitem{virtualzcf}
D.~Moutevelis, J.~Roldán-Pérez, M.~Prodanovic, and F.~Milano, ``Design of
  virtual impedance control loop using the complex frequency approach,'' in
  {\em 2023 IEEE Belgrade PowerTech}, pp.~1--6, 2023.

\bibitem{bernal2024improving}
R.~Bernal and F.~Milano, ``Improving voltage and frequency control of ders
  through dynamic power compensation,'' {\em Electric Power Systems Research},
  vol.~235, p.~110768, 2024.

\bibitem{hillstabilitysync}
L.~Zhu and D.~J. Hill, ``Stability analysis of power systems: A network
  synchronization perspective,'' {\em SIAM Journal on Control and
  Optimization}, vol.~56, no.~3, pp.~1640--1664, 2018.

\bibitem{motter2013spontaneous}
A.~E. Motter, S.~A. Myers, M.~Anghel, and T.~Nishikawa, ``Spontaneous synchrony
  in power-grid networks,'' {\em Nature Physics}, vol.~9, no.~3, pp.~191--197,
  2013.

\bibitem{milano2020frequency}
F.~Milano and A.~Manjavacas, {\em Frequency Variations in Power Systems:
  Modeling, State Estimation, and Control}.
\newblock IEEE Press, Wiley, 2020.

\bibitem{milanoscripting}
F.~Milano, {\em Power system modelling and scripting}.
\newblock Springer Science \& Business Media, 2010.

\bibitem{WECCGFM}
W.~Du, ``Model specification of droop-controlled, grid-forming inverters
  (regfm\_a1),'' 9 2023.

\bibitem{dome}
F.~Milano, ``A {Python}-based software tool for power system analysis,'' in
  {\em IEEE PES General Meeting}, pp.~1--5, IEEE, 2013.

\bibitem{kundur}
P.~Kundur, N.~Balu, and M.~Lauby, {\em Power System Stability and Control}.
\newblock EPRI power system engineering series, McGraw-Hill Education, 1994.

\bibitem{bizarri}
F.~Bizzarri, A.~Brambilla, and F.~Milano, ``The probe-insertion technique for
  the detection of limit cycles in power systems,'' {\em IEEE Transactions on
  Circuits and Systems I: Regular Papers}, vol.~63, no.~2, pp.~312--321, 2016.

\end{thebibliography}
\end{document}